# AN ENHANCED DEEP LEARNING TECHNIQUE FOR PROSTATE CANCER IDENTIFICATION BASED ON MRI SCANS

Hussein Hashem[1], Yasmin Alsakar[2], Ahmed Elgarayhi[1], Mohammed Elmogy[2,*], and Mohammed Sallah[1,3]

[1] Applied Mathematical Physics Research Group, Physics Department, Faculty of Science, Mansoura University, Mansoura 35516, Egypt

[2] Information Technology Department, Faculty of Computers and Information, Mansoura University, Mansoura 35516, Egypt

[3] Higher Institute of Engineering and Technology, New Damietta City, Egypt

Corresponding author: Mohammed Elmogy (melmogy@mans.edu.eg)

The first two authors contributed equally to this work and shared the first authorship.

The fourth and fifth authors are sharing the senior authorship.

## ABSTRACT

Prostate cancer is the most dangerous cancer diagnosed in men worldwide. Prostate diagnosis has been affected by many factors, such as lesion complexity, observer visibility, and variability. Many techniques based on Magnetic Resonance Imaging (MRI) have been used for prostate cancer identification and classification in the last few decades. Developing these techniques is crucial and has a great medical effect because they improve the treatment benefits and the chance of patients' survival. A new technique that depends on MRI has been proposed to improve the diagnosis. This technique consists of two stages. First, the MRI images have been preprocessed to make the medical image more suitable for the detection step. Second, prostate cancer identification has been performed based on a pre-trained deep learning model, InceptionResNetV2, that has many advantages and achieves effective results. In this paper, the InceptionResNetV2 deep learning model used for this purpose has average accuracy equals to 89.20%, and the area under the curve (AUC) equals to 93.6%. The experimental results of this proposed new deep learning technique represent promising and effective results compared to other previous techniques.

## 1. INTRODUCTION

The prostate is a muscle-driven mechanical switch between urine and ejaculation and an auxiliary gland of the male reproductive system [1]. Prostate cancer (PCa) significantly affects societies worldwide and is currently the most prevalent cancer among males. It refers to the uncontrolled growth of cells in the prostate. Prostate glands (where is the PCa) in young men are small. As one gets older, the prostate gland gets bigger. It can restrict urinary flow between the bladder and the urethra as it gets bigger. The main challenge of diagnosing PCa is that most men with PCa will have no symptoms, especially at the early stages of PCa [2].

Moreover, when PCa advances to higher stages, it can cause men to have poor flow and incomplete emptying of urine, which are the same symptoms of prostate enlargement. In its higher stages, PCa spreads outside the prostate gland. This spread outside the prostate gland has been confirmed to be an important prognostic factor for recurrence. Therefore, early diagnosis of PCa is challenging but is crucial as it enables physicians to treat PCa before it develops into a fatal disease [3].

Urologists recommend starting to screen men at the age of 40. The current two initial techniques for early diagnosis of PCa are a digital rectal examination (DRE) and prostate-specific antigen (PSA) blood screening. DRE is a physical examination in which a physician manually checks the prostate gland through the rectum. This technique is highly invasive, and its accuracy is low. PSA screening evaluates the concentration of PSA in the blood as the PCa indicator. PSA screening has high false positive rates as elevated levels of PSA in the blood can be caused due to other diseases such as prostate enlargement or inflammation [4]. If either DRE or PSA is abnormal, the physician recommends a prostate biopsy, the definitive technique for diagnosing PCa. However, the biopsy is an expensive and painful technique that can miss about 35% of clinically significant PCa. The most common surgical operation performed in men is surgery for benign prostatic hyperplasia, which costs more than a billion dollars annually.

The difficulty of diagnosing PCa comes from both non-aggressive PCa's and aggressive PCA's at their early stages have no symptoms. A combination of the DRE, PSA test, and transrectal ultrasound (TRUS)-guided needle biopsy are used to diagnose prostate cancer [5]. Prostate cancer diagnosis by MRI will be important when there is a great contradiction between the PSA level and the outcomes of the TRUS-guided biopsy, which allows physicians to visualize the shape of the prostate and identify the suspicious areas within the prostate gland [6].

PCa detection and localization from MRI data is not easy [7]. Therefore, building CAD systems for identifying PCa is still a work in progress [8]. The accuracy, speed, and automation of these CAD systems vary. However, some processing processes, including prostate segmentation, diagnostic characteristics extraction, and deep learning classification, are applied to diagnose PCa [9].

The main contribution of this paper is to enhance the accuracy and AUC of techniques used to diagnose and detect prostate cancer disease. Firstly, medical image preprocessing has been performed to improve the MRI image problems. Secondly, three pre-trained models have been applied for prostate cancer identification, and the most effective model is the InceptionResNetV2.

The following paper structure is outlined as follows. Section 2 presents the related work on prostate cancer identification. Section 3 introduces the material and proposed method used. Section 4 presents the results and discussion. Section 5 discusses comparison and discussion. Finally, the conclusion and future directions have been presented in Section 6.

## 2. RELATED WORK

Many important research improvements have been made around prostate cancer detection and identification. This section introduces a summary of previous related research on this topic. Table 1 summarizes the method used, strengths, and limitations of previously discussed studies.

Sun et al. [10] used multi-parametric MRI (mpMRI) data and applied the support vector machine (SVM) to detect the prostate tumor site. Before radical prostatectomy, they recorded 16 patients with in vivo MP-MRI data. The sequences used were T2-weighted, diffusion-weighted, and dynamic contrast-enhanced imaging. A Gaussian kernel SVM was trained and tested on distinct patient data subsets. Leave-one-out cross-validation was applied to optimize the parameters. The prediction accuracy was between 70.4 and 87.1 %, and the AUC of the ROC was between 0.81 and 0.94 %.

Bhattacharya et al. [11] employed MRI to diagnose PCa in an automated way. They were trained and verified using a 98 men dataset, comprising 74 men with radical prostatectomy and 24 with normal prostate MRI. CorrSigNIA was tested on three different groups: 55 men with radical prostatectomy, 275 men with targeted biopsies, and 15 men with a normal prostate MRI. CorrSigNIA had an accuracy of 80% in separating men with and without cancer, with a lesion-level ROC-AUC of 0.81±0.31.

Polymeria et al. [12] created an AI-based method to autonomously estimate the prostate and its tumor in 145 patients' PET/CT scans. Between April 2008 and July 2015, the algorithm was tested on 285 high-risk patients who were investigated with 18F-choline PET/CT for primary staging. The prostate gland's tumor fraction and the whole tumor's lesion uptake were obtained automatically. A Cox proportional-hazards regression model was applied to present the relationships between these measures, age, PSA, Gleason score, and PCa-specific survival. They discovered that total prostate tumor volume (p ¼ 0.008), tumor fraction of the gland (p ¼ 0.005), total prostate lesion uptake (p ¼ 0.02), and age (p ¼ 0.01) were all linked with disease-specific survival, whereas SUVmax (p ¼ 0.2), PSA (p ¼ 0.2), and Gleason score (p ¼ 0.8) were not.

Bleker et al. [13] used MRI scans to develop a radionics technique for extracting characteristics from an auto-fixed volume of interest (VOI). They relied on 206 patients with 262 PZ lesions identified using the mpMRI prostate imaging reporting and data system. CS PCa was characterized as Gleason scores greater than 6. To extract features, an auto-fixed 12-mm spherical VOI was placed around a pinpoint in each lesion. Compared to other techniques, dynamic contrast-enhanced imaging (DCE), multivariate feature selection, and extreme gradient boosting (XGB) are important. They discovered that the best model with an (AUC) of 0.870 (95% CI 0.980–0.754) with characteristics from T2-weighted (T2-W) + diffusion-weighted imaging (DWI) + DCE.

Castillo et al. [14] used computer-aided prostate analysis to improve the diagnosis of severe PCa that relies on multi-parametric MRI (mpMRI). For significant-PCa segmentation and/or classification, they used deep-learning and radiomics-based algorithms. Two consecutive patient cohorts (371 individuals) from their center were used in this investigation, along with two other datasets. One of these external sets was a publically available patient cohort (195 patients), while the other contained information from patients from two hospitals (79 patients). One of their patient cohorts (271 patients) was used for deep learning and radiomics model construction throughout training, while the other three (374 patients) were preserved as unseen test sets. The area under the receiver-operating-characteristic curve was used to evaluate the models' performance (AUC). While the deep learning approach, AUCs achieved 0.70, 0.73, and 0.44.

Giannini et al. [15] evaluated non-experienced readers' diagnostic skills while reporting using a CAD system's likelihood map using multi-parametric MRI. The study involved 90 patients (45 with at least one clinically significant biopsy-confirmed PCa). Patients who had at least one clinically significant lesion (GS > 6) exhibited substantially better sensitivity in the CAD-assisted mode (68.7% vs. 78.1%) (p = 0.018).

Yuan et al. [16] trained a deep convolutional neural network with three branch methods to detect discriminative features in prostate images and aid doctors in automatically detecting PCa. They developed a multi-parametric magnetic resonance transfer learning (MPTL) technique for autonomously staging PCa. They first created a deep neural network with three branches. They then used the model to calculate multi-parametric MRI images (mp-MRI) features such as T2w transaxial, T2w sagittal, and obvious diffusion coefficient (ADC). To assess the validity and effectiveness of the proposed MPTL model, they tested 132 cases from our institutional review board-approved patient dataset and 112 instances from the PROSTATEx-2 challenge. Their model had a high accuracy of 86.92% for PCa classification.

Sushentsev et al. [17] evaluated the effective score system against multiple MRI-derived delta-radiomics models (AS). Patients with AS who had biopsy-proven PCa and had at least one repeat-targeted biopsy were included in the study. The parenclitic networks (PN), least absolute shrinkage and selection operator (LASSO) logistic regression, and random forests (RF) methods were used to predict T2WI- and ADC-derived delta-radiomics characteristics using baseline and latest accessible MRI data. The AUCs were compared using DeLong's test, and standard discrimination measures and areas under the ROC curve (AUCs) were determined. Their study included 64 patients (27 progresses and 37 non-progresses). The AUC for PRECISE (84.4%) was not significantly higher than the AUCs for PN, LASSO regression, and RF, which were 81.5 percent, 78.0 percent, and 80.9 percent, respectively (p = 0.64, 0.43, and 0.57, respectively).

**Table 1: The comparison between previous methods.**

| References | Method | Strengths | Limitations |
|---|---|---|---|
| Sun et al. [4] | The Gaussian kernel SVM for detecting prostate tumour location. | An effective method for detection. | Low accuracy in prostate cancer detection. |
| Bhattacharya et al. [5] | Radiology-pathology fusion-based algorithm. | The reduction in biopsy samples used for aggressive cancer detection. | Low accuracy and AUC. |
| Polymeria et al. [6] | AI-based algorithm was used. | AI-based measurements were very important in prostate tumor detection. | Less medical images effected on results. |
| Bleker et al. [7] | Dynamic contrast-enhanced feature method and multivariate feature selection and extreme gradient boosting (XGB) was used. | Effective method in prostate cancer diagnosis. | Low accuracy and AUC. |
| Castillo et al. [8] | Radiomics model and deep learning. | Significant prostate cancer identification. | Low accuracy and AUC. |
| Giannini et al. [9] | Computer Aided Diagnosis and image Interpretation | Increases the per-patient sensitivity in prostate cancer identification. | Needed dataset with specific conditions. |
| Yuan et al. [10] | Deep CNN and three technologies that were T2w transaxial, T2w sagittal, and apparent diffusion coefficient (ADC) features used. | learned discriminative features in prostate images and classified the cancer accurately. | Low accuracy in detection. |
| Sushentsev et al. [11] | The parenclitic networks (PN), least absolute shrinkage and selection operator (LASSO) logistic regression, and random forests (RF) methods were used. | Effective method in detecting prostate cancer. | Low AUC in detection. |

## 3. MATERIAL AND METHODS

The proposed method for prostate cancer identification has been performed in two stages on MRI images [18], [19] as these types of images have evolved their capabilities for detecting prostate cancer [20], [21]. This proposed method has been dependent on deep learning. Although many systems have generated satisfactory results, these systems depend on handcrafted features. There is an alternative system that automatically learns the features. Deep learning structures, especially convolutional neural networks (CNN), automatically learn multiple levels of features from data in a hierarchical way. These structures have produced precise results in various computer vision tasks including lesions detection tasks. Figure.1 illustrates the stage of our proposed method.

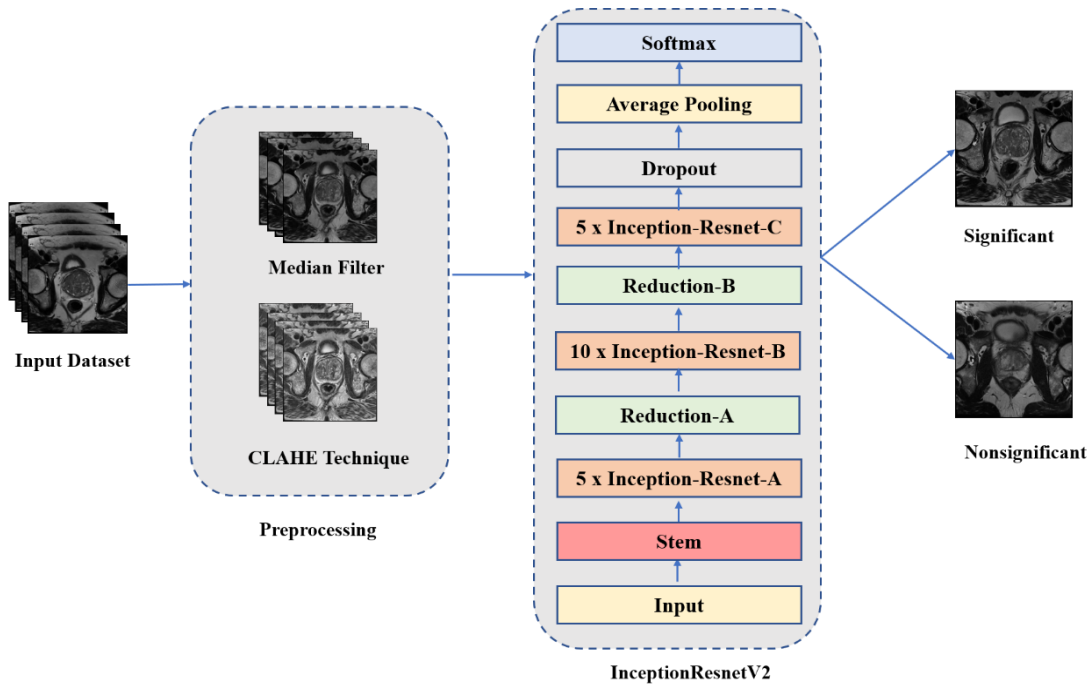

**Figure 1: Prostate cancer identification framework based on InceptionResnetV2.**

### 3.1 Preprocessing

For image preprocessing [22], some techniques are applied to the dataset for image enhancement. This image enhancement technology makes the images more suitable than the original image. The steps in the preprocessing techniques included noise removal and contrast enhancement.

- **Noise Removal**

    Image noise is the arbitrary variation in color information or brightness in images produced by medical scanners or devices. Image noise is considered an undesirable thing during the acquisition of an image. There are many reasons for noise in medical images. medical images include some visual

noise. This noise presence makes images more grainy, mottled, textured, or snowy. Medical images have many types of noise, such as salt and pepper and Gaussian noise.

So, a filtering technique has been used to remove noise from images for solving noise removal. A median filter has been used for smoothing the non-repulsive noise without preserved images and blurring edges. This makes the images more suitable for MRI image enhancement. The median filter applies spatial processing to determine which pixels in medical images have been affected by impulse noise. The median filter classifies pixels as noise by comparing them with their neighboring pixels. This pixel that is different from its neighbor is labeled as an impulse. The median pixel values are used to replace the noise pixels.

- **Contrast enhancement**

  Contrast Limited Adaptive Histogram Equalization is a version of the AHE technique. CLAHE is used to solve the problems of AHE. The AHE causes noise over-amplification, so the CLAHE limits this noise. The CLAHE divides the medical image into multi subblocks and then performs the histogram equalization for each part of the whole image. Figure.2 indicates the impact of preprocessing techniques on MRI prostate cancer images.

### 3.2 Prostate cancer identification

Creating a deep neural network from the start to the end is challenging. Weights are randomly assigned before the training process, and after that, they are altered repeatedly, which depends on the used datasets and the loss function in a large deep neural network. The weight change process takes a long time, and because of training data paucity, the deep network may also become overfit.

There are many advantages to applying CNN over traditional neural networks. Firstly, CNN [23], [24] contains various numbers of layers than traditional neural networks. The layers increase allows CNN to learn high levels of abstraction because the first layers learn primitive components, and the last uses these trained and learned primitive features to generate high-level features. CNNs are used to automatically learn the features. Secondly, CNN takes 2D images and 3D volumes directly as inputs and doesn't need to convert them into vectors. Thirdly, the connections of the network and the CNNs parameters are fewer than in the traditional neural network.

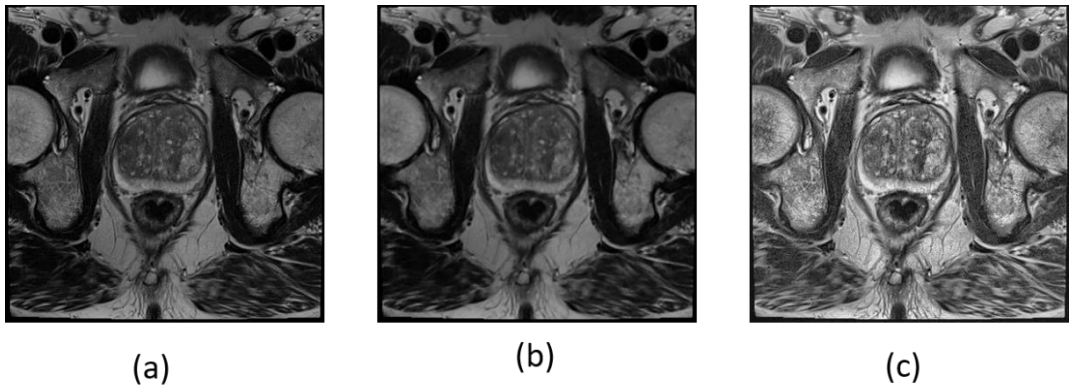

**Figure 2: MRI medical images preprocessing. (a) Original image, (b) Image after applying a median filter, and (c) Image after applying CLAHE technique.**

Transfer learning [25], [26] is an important tool that uses a convolutional deep neural network model that has been previously trained using a variety of datasets. Many models use transfer learning for object recognition and other related computer vision tasks, such as InceptionV3and ResNet50 models. This paper applies three different CNN models for prostate cancer identification.

- **Inception-ResNet-v2**

    Inception-ResNet-v2 [27] is a fusion of two recent networks that are a recent version of Inception architecture and residual connections. These models are essential for their multi-branch architecture. They have some filters that are concatenated together in each branch. The split-transform-merge model of the inception is observed as an essential representational ability in its dense layers. The Inception-ResNet-v2 uses residual connections with better efficiency.

    Inception-ResNet-v2[27], [28] only computes batch-normalization on the first traditional layers but not on the summations. This simplification has been applied to reduce the consumed memory footprint and increase the number of potential inception blocks. Inception-ResNet-v2 is used to stabilize the learning and training by scaling down the residuals before adding them to previous activation layers.

### 4. RESULTS AND DISCUSSION

To evaluate the proposed scheme performance, python 3.7.6 and PyCharm 2019.3.3 with NumPy, Keras, SkLearn, and matplotlib libraries have been used for testing and validating several experiments. Our system was run on a machine of core i7/4.5. It has 16 GB RAM and an NVIDIA GeForce GTX with 4 GB VRAM. The next subsections describe the tested dataset and the standard evaluation metrics. Finally, comparisons and discussion are proposed.

## 4.1 Dataset Description

The dataset [29] consists of 1528 prostate MRI images in the transverse plane. PROSTATEx Dataset and Documentation provided the images and classification. This dataset contains 64 patients' MRI images. These patients should have a single prostate MRI finding for more accurate training. This dataset aims to train and learn a convolutional neural network and other models for classifying new images into clinically significant and nonsignificant.

## 4.2 Performance evaluation metrics

Some evaluation metrics have been used for testing the classifier's performance. Mathematical formulations have been presented to compute the evaluation metrics. TP is the true positive value number that indicates the correct classification of a significant person. FP is the false positive number that indicates the false classification of a significant person. TN is the true negative value number that indicates the correct classification of a nonsignificant person. FN is the false negative number that indicates the false classification of a nonsignificant person. The significant person that has prostate cancer and nonsignificant hasn't prostate cancer.

Some metrics, such as accuracy, have been used to evaluate this system. Sensitivity is the true positive rate (TPR) which indicates positively tested subjects through the examination. Specificity is the true negative rate (TNR) which indicates negative tests identified correctly, F1-score, Matthews correlation coefficient (MCC), and area under the curve (AUC).

MCC [30] was proposed in 1975 for chemical structure analysis and was developed again for machine learning. When the used dataset is unbalanced or if one class is more extensive than the other, the accuracy measure cannot be considered reliable. The solution to this problem is the MCC. AUC is a standard and essential method used in the classifier's evaluation. The AUC is a part of a square unit. AUC is a part of a square area whose value ranges from 0 to 1. Any classifier's value should be bigger than 0.5. ROC is an essential and great test plotted based on specificity and sensitivity.

$$Accuracy = \frac{TP+TN}{TP+FP+TN+FN} \quad (1)$$

$$Precision = \frac{TP}{TP+FP} \quad (2)$$

$$Recall = Sensitivity = \frac{TP}{TP+FN} \quad (3)$$

$$\text{Specificity} = \frac{TN}{FP+TN} \quad (4)$$

$$F1 - \text{score} = 2 \times \frac{(recall \times precision)}{(recall + precision)} \quad (5)$$

$$MCC = \frac{(TP*TN)-(FP*FN)}{\sqrt{(TP+FP).(TP+FN).(TN+FP).(TN+FN)}} \quad (6)$$

$$AUC = \int_0^1 TPR \, d(FPR) \quad (7)$$

## 5. COMPARISONS AND DISCUSSION

In this section, the experimental comparisons have been presented that show the results. Two models have been trained on the dataset for prostate cancer identification, such as InceptionV3 and ResNet50. As we mentioned in the proposed method, InceptionResnetV2 has higher accuracy.

- **Inception-V3 Model**

    The inceptionV3 model is a new and edited version of the inception-V1 model. For more model adaption, this model applies many numbers of approaches for network optimization. This model has more extensive than the Inception-V1 and Inception-V2 models. The Inception-V3 is a deep CNN learned and trained directly on a low-configuration computer. It is fairly challenging to train and takes a long time, and this problem is solved through transfer learning that saves the model's last layer for new categories. The parameters of Inception-V3 of the previous layers are kept, and Inception-V3 is constructed as the final layer is eliminated through the transfer learning technique.

    The Inception-V3 model [31] consists of AvgPool, MaxPool, Convolution, Concat Layer, Fully Connected layer, Dropout, and Softmax Function. The CNN characterization is provided through the sharing and connectivity of the weights, where it computes the outcome of neurons related to present local regions from the previous layer. Additionally, it shares the neuron weights and corresponds to the kernels at the same layers. In the classification part, the fully connected layers take the feature outputs of Inception-V3 and custom-generated segmented features.

- **ResNet50 Model**

    ResNet50 [32] is a CNN model that has 50 layers. The ImageNet database, which has been learned and trained on many images, makes a pre-trained version of the network that can be imported. ResNet50 is suggested as a solution for the challenges of training the CNN model. The ResNet50 advantage is that this model's performance doesn't decrease even though the architecture is getting deeper. The computation is made lighter, and the

ability of the training network is better. The ResNet50 model is made by skipping connections on two to three layers that contain ReLU and batch normalization among the architectures.

Table 2 and Figure 3 show the comparisons and performance results between the three models used for prostate cancer identification. InceptionResnetV2 has the highest results, indicating that this model can detect and identify prostate cancer and classify the MRI images as significant or nonsignificant. The overall accuracy, precision, sensitivity, specificity, F1-score, MCC, and AUC of the proposed method based on InceptionResnetV2 are 89.2%, 93.5%, 84.2%, 94.2%, 88.6%, 78.7% and 93.6 respectively which shows superiority compared with other models.

**Table 2: The performance evaluation of the proposed method and other used models for comparisons.**

| Network | Accuracy | Precision | Sensitivity | Specificity | F1-score | MCC | AUC |
|---|---|---|---|---|---|---|---|
| **InceptionV3** | 87 | 92.4 | 80 | 93.3 | 84.7 | 74.8 | 89.0 |
| **Resnet50** | 76.3 | 89.9 | 59.2 | 93.3 | 71.4 | 56.9 | 84.6 |
| **InceptionResnetV2** | **89.2** | **93.5** | **84.2** | **94.2** | **88.6** | **78.7** | **93.6** |

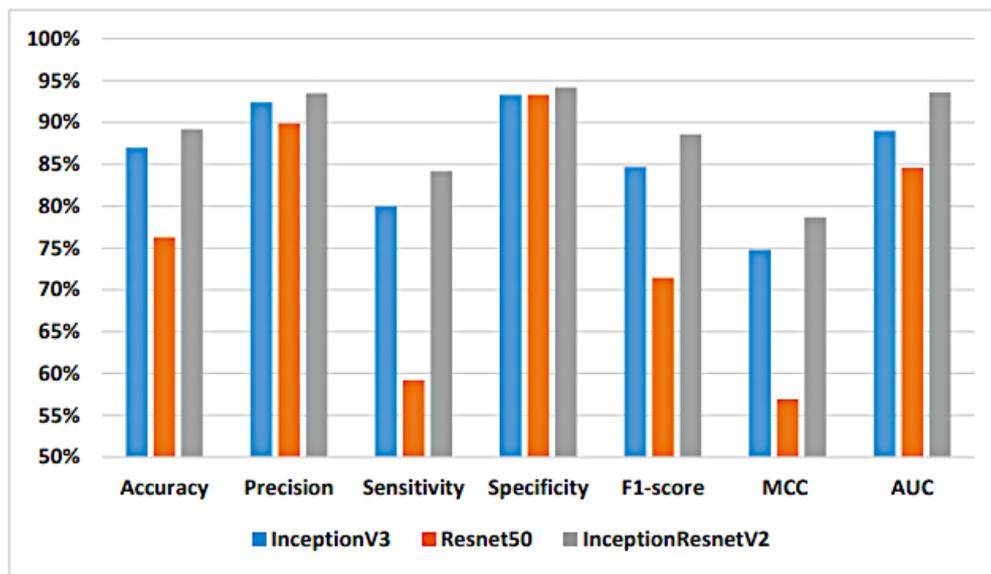

**Figure 3: The comparison results of the proposed method and other used models.**

Figure 4 shows the ROC curve between FPR and TPR for the three deep learning models used for identifying and detecting prostate cancer. After applying these three models, the InceptionResNetV2 is superior to the two other deep learning models.

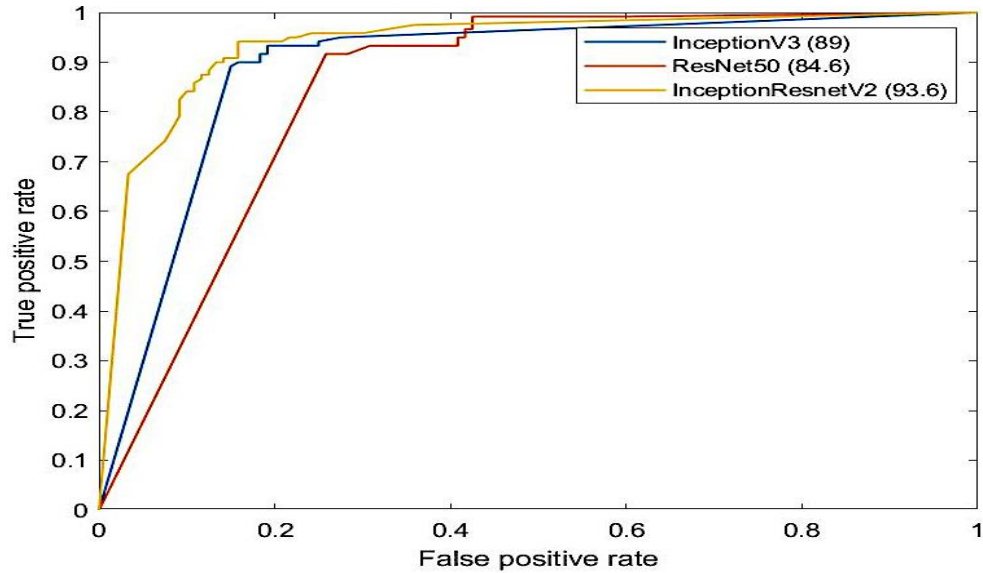

**Figure 4: AUC for three models used for prostate cancer identification.**

Comparison with the previous research is proposed in order to compare the proposed scheme performance with different methods on the same dataset. Table 3 and Figure 5 indicate the comparison that shows superiority compared with the state-of-the-art with AUC, which was 936%.

**Table 3: Performance comparison between the proposed approach and other related methods based on AUC.**

| Methods | AUC (%) |
|---|---|
| Grebenisan et al. [33] | 75.0 |
| Sobecki et al. [34] | 84.0 |
| Pin et al. [35] | 73.0 |
| Proposed Method | **93.6** |

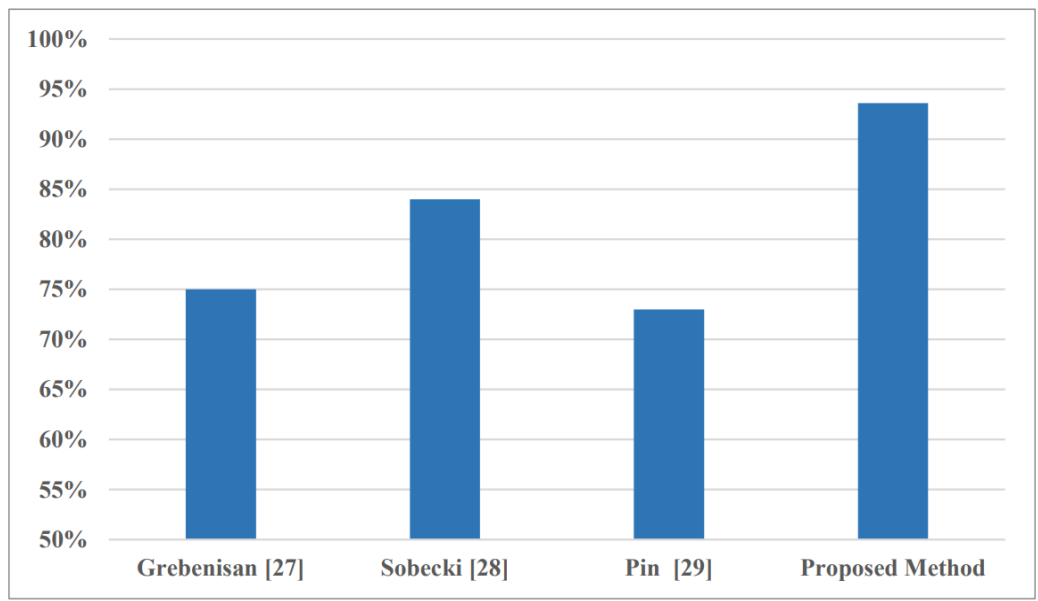

**Figure 5: Performance comparison results between the proposed approach and other related methods based on AUC.**

## 6. CONCLUSIONS

This paper presented a new technique for prostate cancer detection and classification into two classes significant and nonsignificant. This technique consists of two stages: preprocessing step for MRI images and prostate cancer identification. It has depended on MRI scans that have been developed for diagnosis improvement. Three different models have been trained for prostate cancer classification. The best is InceptionResNetV2, which has 89.2% accuracy and 93.6% performance, representing promising results of the proposed new deep learning technique. In future work, we will try to apply more deep learning techniques to increase the accuracy of the proposed system. In addition, we will use multi-modalities to evaluate the PC.